\documentclass[]{article}

\usepackage{arxiv}

\usepackage[utf8]{inputenc} 
\usepackage[T1]{fontenc}    
\usepackage{hyperref}       
\usepackage{url}            
\usepackage{booktabs}       
\usepackage{amsfonts}       
\usepackage{nicefrac}       
\usepackage{microtype}      
\usepackage{lipsum}
\usepackage{graphicx}
\usepackage{tikz}
\usepackage{tabularx} 
\usepackage{subcaption}
\usepackage{svg}            
\usepackage{layouts}        
\usepackage{comment}        
\usepackage{siunitx}

\newcommand{\DTA}{\mbox{``Deutschlandticket''}~}
\newcommand{\NET}{\mbox{``9-Euro-Ticket''}~}

\title{A nation-wide experiment, part II: the introduction of a 49-Euro-per-month travel pass in Germany -- An empirical study on this fare innovation}

\author{
Allister Loder \\
Technical University of Munich\\
TUM School of Engineering and Design\\
Chair of Traffic Engineering and Control\\
Arcisstrasse 21, 80333 Munich \\
\texttt{allister.loder@tum.de}\\
\And
Fabienne Cantner\\
Technical University of Munich\\
TUM School of Management\\
TUMCS for Biotechnology \& Sustainability \\ 
Am Essigberg 3, 94315 Straubing\\
\texttt{fabienne.cantner@tum.de}\\
\And
Lennart Adenaw\\
Technical University of Munich\\
TUM School of Engineering and Design\\
Chair of Automotive Technology \\
Boltzmannstrasse 15, 85748 Garching\\
\texttt{lennart.adenaw@tum.de}\\
\And
Markus B. Siewert\\
Munich School of Politics and Public Policy\\
TUM Think Tank\\
Richard-Wagner-Straße 1, 80333 München\\
\texttt{markus.siewert@hfp.tum.de}
\And
Sebastian Goerg\\
Technical University of Munich\\
TUM School of Management\\
TUMCS for Biotechnology \& Sustainability \\ 
Am Essigberg 3, 94315 Straubing\\
\texttt{sebastian.goerg@tum.de}\\
\And
Klaus Bogenberger\\
Technical University of Munich\\
TUM School of Engineering and Design\\
Chair of Traffic Engineering and Control\\
Arcisstrasse 21, 80333 Munich\\
\texttt{klaus.bogenberger@tum.de}\\
}

\begin{document}
\maketitle
\begin{abstract}
In a response to the 2022 cost-of-living crisis in Europe, the German government implemented a three-month fuel excise tax cut and a public transport travel pass for 9 Euro per month valid on all local and regional services. Following this period, a public debate immediately emerged on a successor to the so-called \NET, leading to the political decision of introducing a similar ticket priced at 49 Euro per month in May 2023, the so-called \DTA. We observe this introduction of the new public transport ticket with a sample of 818 participants using a smartphone-based travel diary with passive tracking and a two-wave survey. The sample comprises 510 remaining participants of our initial \NET study from 2022 and 308 participants recruited in March and early April 2023. In this report we report on the status of the panel before the introduction of the \DTA.
\end{abstract}


\section{Introduction}

In a response to the 2022 cost-of-living crisis in Europe, the German government implemented a three-month cut on the fuel excise tax and a public transport travel pass for 9 Euro per month valid on all local and regional services in the summer months of 2022, i.e., June, July, and August. While the reduction in the fuel excise tax has not been experienced by drivers as such due to global market price fluctuations, the so-called \NET impacted travel behavior substantially. This behavioral intervention, which could be one of the largest public transport pricing natural travel behavior experiments, has been studied by several studies. All of them reported a substantial increase in public transport usage during the validity period of the \NET \cite{deutsche_bahn_ag_abschlussbericht_2022,gaus_9-euro-ticket_2023,nationwide_report4,nationwide_report5,nationwide_trb2023}.

During the \NET validity period, the study of the Association of German Transport Companies surveyed more than 200,000 people in Germany \cite{deutsche_bahn_ag_abschlussbericht_2022}. The study reports that around 20~\% of all \NET customers were new customers to public transport. Out of all public transport trips in the months of June, July and August 2022, 17~\% of trips have been shifted from other transport modes and 10~\% of trips have been shifted from the car to public transport, in the countryside even 13 to 16~\%. Importantly, 16~\% of all trips  would not have taken place at all without the \NET, i.e., their correspond to induced demand. In addition, trip distances increased by 38~\% in the \NET period. Nevertheless, car usage numbers returned to pre-\NET levels in the month after the ticket's  \cite{deutsche_bahn_ag_abschlussbericht_2022}. In a Munich-oriented study \cite{nationwide_trb2023}, a modal shift from car to public transport in the order of five percent of the average daily travel distance was observed \cite{nationwide_report5}. Around half the sample did not report any substantial change in car use, while only around 8~\% of the sample reported less car and more public transport usage instead. The study also suggests that an activation effect of the \NET exists because around four percent of the sample who did not previously used public transport frequently are now using it on a regular basis. Interestingly, the share of substituted car trips after the 9-Euro-Ticket period did not return to zero levels: around ten percent of the sample still substitutes some car trips. For a nation-wide sample, similar findings were made by \cite{gaus_9-euro-ticket_2023} who state that the ticket did not lead to a shift in daily mobility, but rather increased leisure travel at the beginning and the end of the ticket's validity period, leaving monetary savings as the main effect of the \NET.

With the introduction of the \DTA as the successor ticket in May 2023, priced at 49~Euro per month , the research question becomes how travel behavior will be affected by this similar, but not identical offering. It differs from the \NET in so far as it costs 49 Euro instead of 9 Euro per month, it is not limited to three months, and it is a subscription instead of a monthly ticket. 

As with the \NET, it can be expected that various research institutions and companies will perform (market) research on this fare innovation and disruption to the German transportation system. We also decided to continue our \NET study ``Mobilit\"{a}.Leben'' \cite{nationwide_report1} to cover the first months of the validity period of the successor ticket. In the sequel of our initial study, some study participants decided to continue their participation, while other participants entered the study after the \NET period. 

In this report, we detail on the panel's composition and its recruiting in Section \ref{sec:recruit}; report on the most recent study activity in Section \ref{sec:activity}; present the panel's travel pass ownership before the introduction of the successor ticket and the panel's current intention to subscribe to this ticket in Section \ref{sec:ticket}. Last, we provide an outlook of our study in Section \ref{sec:outlook}.

\section{Panel and recruiting} \label{sec:recruit}

The study to observe the travel behavior impacts of the so-called ``49-Euro-Ticket'' or \DTA (Germany-Ticket) uses a panel that comprises study participants from our previous \NET study \cite{nationwide_report1} as well as participants recruited in the period before the start of \DTA in May 2023. Table \ref{tab:panel_composition} shows the time of recruitment. It can be seen that around two-thirds of the panel result from the initial \NET study, while one-third has been recruited in the months before the start of the \DTA.

The recruitment for the initial \NET study as well as the sequel study on the \DTA was based on self-enrollment. The study was advertised on social media, in TV shows, and in newspapers. Self-enrollment was possible until April 21, 2023, to ensure that at least one week of travel can be observed using the smartphone-based travel diary. 

\begin{table}[htbp]
    \centering
    \begin{tabularx}{\textwidth}{X r r}
        \toprule
        Recruited through & N & Share of participants  \\ 
        \midrule
        9-Euro-Ticket panel & 510 & 62.35~\% \\
        Until February 2023 & 32 & 3.91~\% \\
        March 2023 & 110 & 13.45~\%  \\
        April 2023 & 166 & 20.29~\% \\
        \midrule
        Total & 818 & 100.00~\%   \\
        \midrule
    \end{tabularx}
    \caption{Panel composition}
    \label{tab:panel_composition}
\end{table}

Table \ref{panel:socio} shows the socio-demographic characteristics of the panel (left column) compared to 2022 German census data\footnote{\url{https://www-genesis.destatis.de/genesis}}. Importantly, our study only recruited participants of 18 years of age and older. In all other age groups except for 75 years and older, our panel is well represented, while the age groups 25-30 and 55-60 seem to be over-represented. Considering gender, we see that compared to census data, males are slightly over-represented. Nevertheless, it can be concluded that our panel provides sufficient variation of the German population in terms of age and gender to derive implications with some degree of external validity.

\begin{table}[htbp]
    \centering
    \begin{tabularx}{\textwidth}{X r r}
        \toprule
 & \multicolumn{1}{l}{Panel} & \multicolumn{1}{l}{Germany census}  \\
        \midrule
less than 15 years & 0.00\% & 14.31\%  \\
15 to 20 years & 2.57\% & 4.71\%  \\
20 to 25 years & 8.07\% & 5.40\%   \\
25 to 30 years & 14.67\% & 5.85\%   \\
30 to 35 years & 9.66\% & 6.82\%   \\
35 to 40 years & 9.90\% & 6.49\%   \\
40 to 45 years & 9.41\% & 6.33\%   \\
45 to 50 years & 7.33\% & 5.86\%   \\
50 to 55 years & 9.78\% & 7.34\%   \\
55 to 60 years & 11.74\% & 8.24\%   \\
60 to 65 years & 6.85\% & 7.26\%   \\
65 to 70 years & 5.01\% & 5.96\%   \\
70 to 75 years & 3.55\% & 5.07\%   \\
75 years and more & 1.47\% & 10.36\%   \\
\midrule
Male & 53.06\% & 49.39\%   \\
Female & 46.09\% & 50.61\%   \\
Diverse & 0.86\% & NA \\
        \midrule
        Total & 100.00~\% & 100.00~\%   \\
        \midrule
    \end{tabularx}
    \caption{Socio-demographic characteristics of the panel.}
    \label{panel:socio}
\end{table}

\section{Study activity}\label{sec:activity}

\begin{figure}
    \centering
    \includegraphics[width=\textwidth]{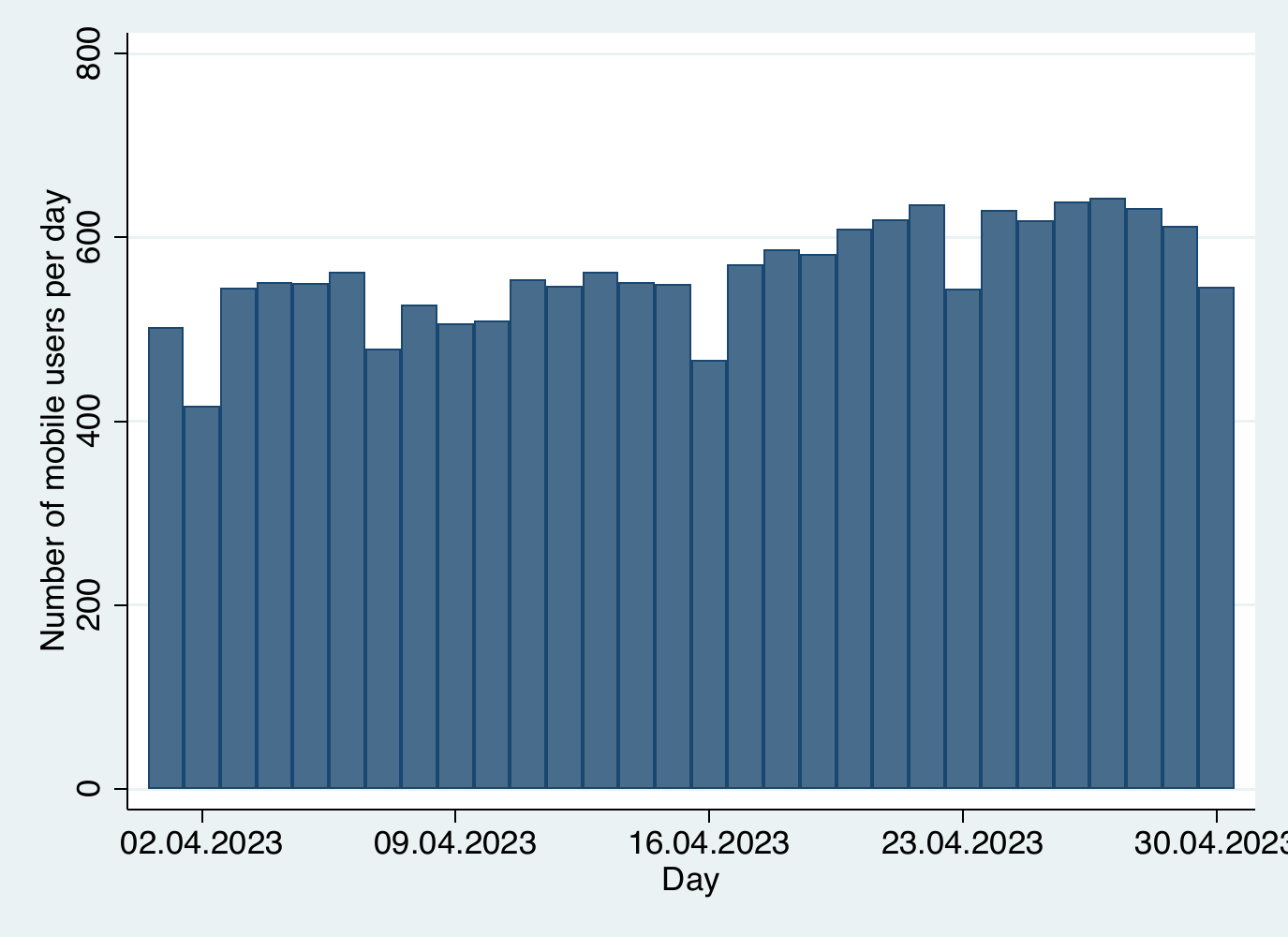}
    \caption{The number of mobile users in the ``Mobilit\"{a}.Leben'' app per day in April 2023.}
    \label{fig:nusers}
\end{figure}

Out of the 818 registered panel participants, 741 did successfully active the smartphone app (around 90~\%), while more than 670 were able to successfully submit travel behavior data (around 83~\%). In other words, around 70 participants who did activate the smartphone app either deleted the app already again or their smartphone  did not report any trips for technical reasons. Considering the panel dynamics, Figure \ref{fig:nusers} shows the number of mobile users in April, i.e., the month before the start of the \DTA. It can be seen that, the number of mobile users is increasing towards the end of the month which can be explained by the fact that recruitment was ongoing. 

\begin{figure}
    \centering
    \includegraphics[width=\textwidth]{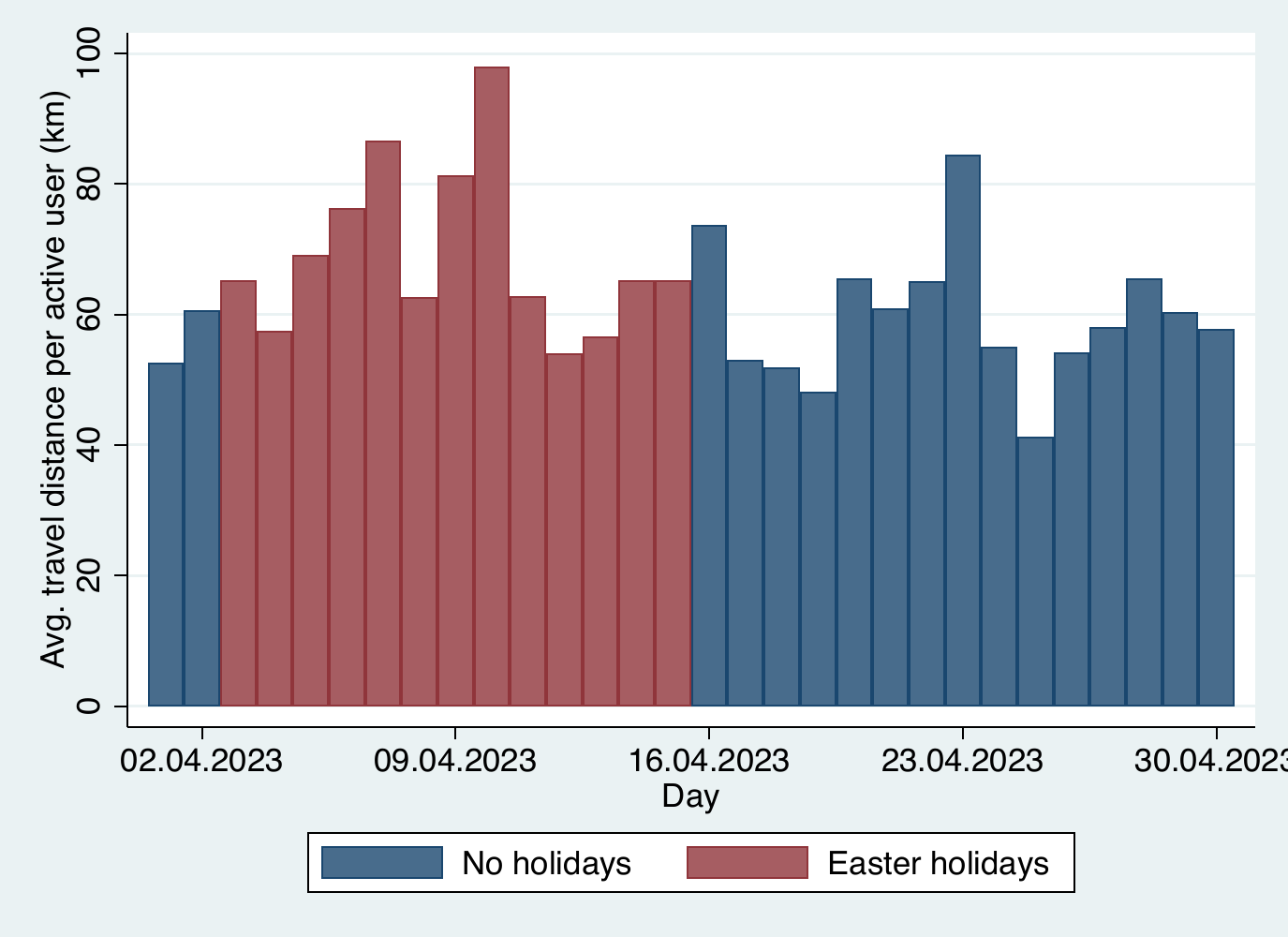}
    \caption{Average daily travel distance of mobile users.}
    \label{fig:daily_dist}
\end{figure}

Considering the travel behavior in April, Figure \ref{fig:daily_dist} shows the average travel distances per day and mobile user. According to the German household travel survey, the average daily travel distance of mobile persons is around 46~kilometers per day \cite{mid_munich2019}. Consequently, the recruited panel is likely more mobile as only few days show a similar average travel distance, while many days, especially during the Easter holidays show substantially larger daily travel distances.

Overall, the observed modal share based kilometers without air travel in April 2023, i.e., the month before the introduction of the \DTA, is 52.7~\% private transport and 37.8~\% public transport. These values are close to the modal share between the values report for Munich (36~\% public transport, 56~\% private transport) and the greater Munich area (28~\% public transport, 55~\% private transport) in 2017 \cite{mid_munich2019}. This is not surprising as the Munich metropolitan region is the focus region of our study.

The April questionnaire, i.e., the questionnaire distributed right before the start of the \DTA has been completed by 644 or around 80~\% of all registered participants. 632 participants, around 77~\% of all registered participants, successfully activated the smartphone app and completed the questionnaire April.  

\section{Travel pass ownership and intentions} \label{sec:ticket}

In April, 47~\% of respondents of our panel have a monthly travel pass, while 53~\% have not. Note that our study has a focus on the Munich metropolitan region \cite{nationwide_report1}. Out of the non-travel-pass owners, 16~\% already bought the \DTA in April. This increases the total number of monthly travel pass owners in the panel by 18~\%. Further 20~\% of the entire panel consider subscribing to the \DTA. If all of them would eventually subscribe to the \DTA, the total number of travel pass owners in the panel would grow by 62~\% compared to the April levels. 

\section{Outlook} \label{sec:outlook}

The presented study will continue until June 2023. It includes observing the travel behavior of the sample for the first two months of the \DTA using the smartphone app and a questionnaire distributed in June, i.e., in the second month of the \DTA validity period. Considering that two-thirds of the sample did already participate in the \NET study, we will be able to compare and explore behavioral differences between the period of the \NET and the \DTA. 

\section*{Acknowledgement}

Allister Loder acknowledges funding by the Bavarian State Ministry of Science and the Arts in the framework of the bidt Graduate Center for Postdocs. Fabienne Cantner acknowledges funding by the Munich Data Science Institute (MDSI) within the scope of its Seed Fund scheme. The authors would like to thank the TUM Think Tank at the Munich School of Politics and Public Policy led by Urs Gasser for their financial and organizational support and the TUM Board of Management for supporting personally the genesis of the project. The authors thank the company MOTIONTAG for their efforts in producing the app at unprecedented speed. Further, the authors would like thank everyone who supported us in recruiting participants, especially Oliver May-Beckmann and Ulrich Meyer from MCube and TUM, respectively.

\bibliographystyle{unsrt}  
\bibliography{references,zotero}  



\end{document}